\documentclass[%
 reprint,
superscriptaddress,
 amsmath,amssymb,
 aps,
]{revtex4-2}
\usepackage{graphicx}
\usepackage{dcolumn}
\usepackage{bm}
\usepackage{hyperref}
\usepackage{float}
\usepackage{booktabs} 
\usepackage{amsmath}
\usepackage{siunitx}  
\usepackage{placeins}

\newcommand{\tblnote}[2][4.8cm]{\parbox[t]{#1}{\raggedright #2}}

\RequirePackage[normalem]{ulem} 
\RequirePackage{color}\definecolor{RED}{rgb}{1,0,0}\definecolor{BLUE}{rgb}{0,0,1} 
\definecolor{GREEN}{RGB}{0,130,0}

\providecommand{\DIFaddbegin}{} 
\providecommand{\DIFaddend}{} 
\providecommand{\DIFdelbegin}{} 
\providecommand{\DIFdelend}{}

\providecommand{\DIFaddbeginFL}{} 
\providecommand{\DIFaddendFL}{} 
\providecommand{\DIFdelbeginFL}{} 
\providecommand{\DIFdelendFL}{}

\newcommand{\DIFscaledelfig}{0.5}
\RequirePackage{settobox} 
\RequirePackage{letltxmacro} 
\newsavebox{\DIFdelgraphicsbox} 
\newlength{\DIFdelgraphicswidth} 
\newlength{\DIFdelgraphicsheight} 
\LetLtxMacro{\DIFOincludegraphics}{\includegraphics} 
\newcommand{\DIFaddincludegraphics}[2][]{{\color{GREEN}\fbox{\DIFOincludegraphics[#1]{#2}}}} 
\newcommand{\DIFdelincludegraphics}[2][]{%
\sbox{\DIFdelgraphicsbox}{\DIFOincludegraphics[#1]{#2}}%
\settoboxwidth{\DIFdelgraphicswidth}{\DIFdelgraphicsbox} 
\settoboxtotalheight{\DIFdelgraphicsheight}{\DIFdelgraphicsbox} 
\scalebox{\DIFscaledelfig}{%
\parbox[b]{\DIFdelgraphicswidth}{\usebox{\DIFdelgraphicsbox}\\[-\baselineskip] \rule{\DIFdelgraphicswidth}{0em}}\llap{\resizebox{\DIFdelgraphicswidth}{\DIFdelgraphicsheight}{%
\setlength{\unitlength}{\DIFdelgraphicswidth}%
\begin{picture}(1,1)%
\thicklines\linethickness{2pt} 
{\color[rgb]{1,0,0}\put(0,0){\framebox(1,1){}}}%
{\color[rgb]{1,0,0}\put(0,0){\line( 1,1){1}}}%
{\color[rgb]{1,0,0}\put(0,1){\line(1,-1){1}}}%
\end{picture}%
}\hspace*{3pt}}} 
} 
\LetLtxMacro{\DIFOaddbegin}{\DIFaddbegin} 
\LetLtxMacro{\DIFOaddend}{\DIFaddend} 
\LetLtxMacro{\DIFOdelbegin}{\DIFdelbegin} 
\LetLtxMacro{\DIFOdelend}{\DIFdelend} 
\DeclareRobustCommand{\DIFaddbegin}{\DIFOaddbegin \let\includegraphics\DIFaddincludegraphics} 
\DeclareRobustCommand{\DIFaddend}{\DIFOaddend \let\includegraphics\DIFOincludegraphics} 
\DeclareRobustCommand{\DIFdelbegin}{\DIFOdelbegin \let\includegraphics\DIFdelincludegraphics} 
\DeclareRobustCommand{\DIFdelend}{\DIFOaddend \let\includegraphics\DIFOincludegraphics} 
\LetLtxMacro{\DIFOaddbeginFL}{\DIFaddbeginFL} 
\LetLtxMacro{\DIFOaddendFL}{\DIFaddendFL} 
\LetLtxMacro{\DIFOdelbeginFL}{\DIFdelbeginFL} 
\LetLtxMacro{\DIFOdelendFL}{\DIFdelendFL} 
\DeclareRobustCommand{\DIFaddbeginFL}{\DIFOaddbeginFL \let\includegraphics\DIFaddincludegraphics} 
\DeclareRobustCommand{\DIFaddendFL}{\DIFOaddendFL \let\includegraphics\DIFOincludegraphics} 
\DeclareRobustCommand{\DIFdelbeginFL}{\DIFOdelbeginFL \let\includegraphics\DIFdelincludegraphics} 
\DeclareRobustCommand{\DIFdelendFL}{\DIFOaddendFL \let\includegraphics\DIFOincludegraphics} 
\RequirePackage{listings} 
\RequirePackage{color} 
\lstdefinelanguage{DIFcode}{ 
  moredelim=[il][\color{red}\sout]{\%DIF\ <\ }, 
  moredelim=[il][\color{blue}\uwave]{\%DIF\ >\ } 
} 
\lstdefinestyle{DIFverbatimstyle}{ 
	language=DIFcode, 
	basicstyle=\ttfamily, 
	columns=fullflexible, 
	keepspaces=true 
} 
\lstnewenvironment{DIFverbatim}{\lstset{style=DIFverbatimstyle}}{} 
\lstnewenvironment{DIFverbatim*}{\lstset{style=DIFverbatimstyle,showspaces=true}}{} 
\setcitestyle{super,sort&compress}

\hypersetup{hidelinks}
\makeatletter
\renewcommand{\frontmatter@thefootnote}{*}
\makeatother

\begin{document}

\preprint{APS/123-QED}

\title{Inverse Design of the Topology–Bandwidth Tradeoff in Valley Photonic Crystals}

\author{Devansh Satra}
\affiliation{Laboratory of Optics of Quantum Materials, Department of Physics, IIT Bombay, Mumbai - 400076, India}
\author{Abhishek Kumar}%
\email{abhishekkumar@jncasr.ac.in}
\affiliation{School of Advanced Materials (SAMaT), Jawaharlal Nehru Centre for Advanced Scientific Research (JNCASR), Bangalore 560064, India}
\affiliation{New Chemistry Unit (NCU), Jawaharlal Nehru Centre for Advanced Scientific Research (JNCASR), Bangalore 560064, India}
\affiliation{Chemistry and Physics of Materials Unit (CPMU), Jawaharlal Nehru Centre for Advanced Scientific Research (JNCASR), Bangalore 560064, India}

\author{Anshuman Kumar}
\email{anshuman.kumar@iitb.ac.in}
\affiliation{Laboratory of Optics of Quantum Materials, Department of Physics, IIT Bombay, Mumbai - 400076, India}
\affiliation{Centre of Excellence in Quantum Information, Computation, Science and Technology, Indian Institute of Technology Bombay, Powai, Mumbai 400076, India}

\begin{abstract}
\noindent\textbf{Abstract:}
Valley photonic crystals provide a route to broadband domain-wall guiding, but increasing the bulk gap can delocalize Berry curvature and reduce the non-quantized valley-Chern number. We introduce a mixed-integer particle-swarm framework that searches polygonal VPC unit cells and maps the Pareto tradeoff between normalized bandgap and valley topology using the bulk objective $\mathcal{T}=(\Delta f/f_0)^2|C_v|$. The selected designs are then evaluated independently using full-wave transport simulations through disorder-averaged propagation-length extraction in $200a$ domain-wall waveguides with positional disorder $\sigma=0.01a$. Across the studied design ensemble, the extracted propagation length exhibits a positive empirical association with $|C_v|$, providing a direct multi-design connection between the patch-integrated valley-Chern number and quantified disorder-limited transport in this VPC family. The framework therefore treats bandwidth, valley topology, and device robustness as distinct but connected quantities, enabling balanced inverse design rather than optimization of a single transport metric alone.
\end{abstract}

\maketitle

\section*{Introduction}
Dense on-chip photonic systems must route light through compact footprints while maintaining low loss, low crosstalk, and single-mode behavior. Conventional dielectric waveguides \cite{Atakaramians:13} incur reflections at discontinuities, bend loss in tight turns, and performance variations under fabrication tolerances, which motivate topological approaches to photonic routing \cite{Lu2014,Khanikaev2017,RevModPhys.91.015006}. Topological photonic devices have been demonstrated across several symmetry classes, including Chern-type, spin-Hall-like, and valley-Hall platforms, enabling robust waveguides \cite{Shalaev2019,Yang2020,Tan:22,PhysRevResearch.6.L022065,doi:10.1126/science.aaq0327}, delay lines \cite{Hafezi2011}, splitters \cite{Cheng2016,PhysRevApplied.18.044080,Wang2024,PhysRevLett.126.230503}, tapers \cite{Flower2023,https://doi.org/10.1002/adom.202300764}, beamformers \cite{Wang2024}, demultiplexers \cite{Kumar2022}, and resonators \cite{PhysRevB.101.205303,JalaliMehrabad:23}. In the present work, we focus specifically on inversion-broken valley photonic crystals (VPCs) \cite{https://doi.org/10.1002/adpr.202100013,Ma_2016}. In these systems time-reversal symmetry is preserved, the net Chern number vanishes, and the relevant topological response is encoded in valley-contrasting Berry curvature localized near the inequivalent $K$ and $K'$ valleys. Domain-wall kink modes formed between inverted VPC domains can exhibit suppressed backscattering when intervalley mixing remains weak, while retaining compatibility with dielectric photonic-crystal fabrication routes.

\begin{figure*}[t]
 \centering
 \includegraphics[width=1\linewidth]{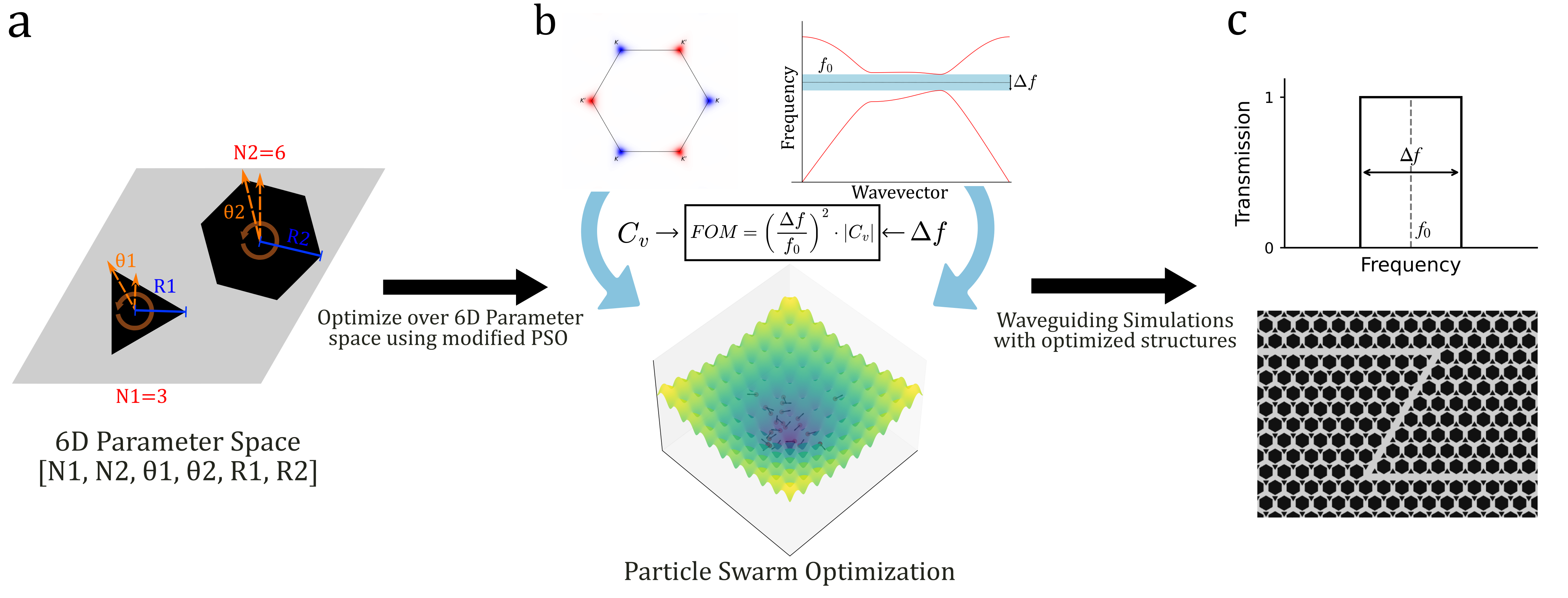}
 \caption{\textbf{Inverse-design workflow for topology-aware bandwidth optimization in valley photonic crystals.}
 \textbf{(a)} Six-dimensional unit-cell parameterization using two symmetry-inequivalent regular polygonal air holes, with continuous variables (feature sizes and rotations) and discrete variables (polygon side counts).
 \textbf{(b)} Optimization objective combines a normalized bulk gap (bandwidth) with a valley-Chern number computed from Berry curvature, forming a single figure of merit explored by mixed-integer PSO.
 \textbf{(c)} Device-level transport diagnostics: optimized structures are interfaced to form domain-wall waveguides and evaluated by full-wave simulations, followed by disorder-averaged propagation-length extraction.}
 \label{fig:1}
\end{figure*}

A useful modern view of topological photonics is that Bloch eigenmodes possess a quantum-geometric structure encoded in the quantum geometric tensor (QGT). For a normalized periodic Bloch eigenstate $\lvert u_n(\mathbf{k})\rangle$, the QGT is
\begin{equation}
Q^{(n)}_{\mu\nu}(\mathbf{k})=
\big\langle \partial_{k_\mu}u_n \big\vert
\Big(1-\lvert u_n\rangle\langle u_n\rvert\Big)
\big\vert \partial_{k_\nu}u_n \big\rangle,
\end{equation}
whose imaginary and real parts define, respectively, the Berry curvature and the quantum metric:
$\Omega_n(\mathbf{k})=2\,\mathrm{Im}\,Q^{(n)}_{xy}(\mathbf{k})$ and
$g^{(n)}_{\mu\nu}(\mathbf{k})=\mathrm{Re}\,Q^{(n)}_{\mu\nu}(\mathbf{k})$.
In VPCs, inversion breaking acts as an effective Dirac mass that gaps the valley degeneracies, produces opposite Berry-curvature hot spots at $K$ and $K'$, and modifies the local mode-overlap geometry. This motivates a design question that is naturally geometric: how can the unit cell be shaped to enlarge the usable valley gap while retaining a strong valley-contrasting Berry-curvature response?

For an interface between two domains with opposite valley topology, the counter-propagating kink modes are separated in momentum space. Their resistance to backscattering is therefore limited by intervalley mixing, which can be induced by disorder or by interface features that provide the required large momentum transfer \cite{Rosiek2023}. Moreover, the valley-Chern number in broad-gap dielectric VPCs is not a strictly quantized invariant equal to $\pm 1/2$ \cite{PhysRevResearch.3.L022025}; as the inversion-breaking perturbation increases, Berry curvature can spread away from the valley patch, causing $|C_v|$ to decrease from its small-gap limit. We therefore treat $|C_v|$ as a quantitative valley-localization descriptor rather than as an exact topological invariant.

Sharp-bend transmission is treated here as a finite-device transport diagnostic rather than as a standalone topological criterion. Interface termination, mode matching, local field chirality, and polarization-singularity structure can all influence bend transmission in photonic-crystal waveguides. In this work, the bulk valley-Chern number is used as a topology descriptor, while FDTD waveguiding simulations and disorder-limited propagation lengths are used as independent device-level transport diagnostics.

Recent inverse-design studies relevant to topological photonics have targeted specific device objectives: prescribed pseudospin functionality in dual-band VPCs \cite{Sato2025InverseVPC}, valley-locked slow-light rainbow trapping \cite{Liu2025LatentDiffusion}, terahertz communication bandwidth and bit-error rate \cite{Li2025THzInverse}, and reduced disorder-induced backscattering at controlled group index \cite{Thompson2026Backscattering}. These studies optimize prescribed functionalities or selected device-level metrics. Here we instead map the compromise between bulk bandwidth and valley topology through $\mathcal{T}=(\Delta f/f_0)^2|C_v|$, and then independently test whether the non-quantized $|C_v|$ component remains empirically associated with disorder-limited propagation length. The resulting framework combines mixed-integer topology–bandwidth Pareto mapping with a cross-design analysis of the empirical relation between the patch-integrated valley-Chern number and disorder-limited propagation length.

Figure~\ref{fig:1} summarizes the workflow. Figure~\ref{fig:1}a defines the six-dimensional mixed discrete--continuous unit-cell parameterization using two symmetry-inequivalent regular polygons. Figure~\ref{fig:1}b shows the bulk objective combining normalized gap and patch-integrated valley-Chern response. Figure~\ref{fig:1}c illustrates the device-level transport diagnostics: optimized bulk designs are interfaced to form domain-wall waveguides, evaluated through straight and bent full-wave simulations, and further tested through the disorder-averaged propagation-length analysis in Fig.~\ref{fig:cv_xi}. Full derivations, convergence tests, parameter bounds, and fit diagnostics are provided in the Supplementary Information.

Ultimately, the goal is to improve the topology--bandwidth compromise, clarify the limits of valley-Chern-based design in broad-gap regimes, and demonstrate the value of explicitly topology-aware bulk objectives for photonic-structure optimization.

\section*{Computational Methodology}

\subsection*{Plane-wave expansion and effective-index model}

We use the plane-wave expansion (PWE) method \cite{joannopoulos2008photonic,sukhoivanov2009photonic} to compute the bulk bands of the inversion-broken VPC unit cell. All calculations reported in this work are performed within a two-dimensional effective-index model. The silicon slab is represented by an in-plane effective refractive index $n_{\mathrm{eff}}=3$ and the air regions by $\epsilon=1$; consequently, both the PWE optimization and the two-dimensional Tidy3D FDTD simulations solve the same TE-like $H_z$-polarized effective-index problem. The present results therefore quantify in-plane band topology, domain-wall confinement, and in-plane scattering. Out-of-plane radiation, substrate leakage, and true three-dimensional slab light-cone effects are not included.

The PWE solver was implemented in-house using a truncated reciprocal-lattice basis. The inverse dielectric function is evaluated from the unit-cell geometry and assembled into the TE $H_z$ eigenvalue problem. The resulting Hermitian matrix is diagonalized for each $\mathbf{k}$ point along the high-symmetry path and on the two-dimensional meshes used for Berry-curvature evaluation. Reciprocal-space cutoffs, Fourier-coefficient evaluation, convergence tests, and matrix-assembly details are given in the Supplementary Information.

\subsection*{Berry curvature and valley-Chern number}

Berry curvature is computed on a discrete $\mathbf{k}$ mesh using the gauge-invariant Fukui--Hatsugai--Suzuki (FHS) link-variable plaquette method \cite{Fukui2005,Wang2020,BlancoDePaz2020}. For the lower band forming the target valley-Hall gap, we define the signed valley-Chern number around the $K$ valley as
\begin{equation}
C_v^{K}=\frac{1}{2\pi}\int_{P_K}\Omega_n(\mathbf{k})\,d^2k,
\label{eq:valley_chern}
\end{equation}
where $P_K$ is the fixed, non-overlapping valley patch shown explicitly in Fig.~\ref{fig:optimized}c and defined geometrically in the Supplementary Information. The corresponding $K'$ value satisfies $C_v^{K'}\simeq -C_v^K$ under time-reversal symmetry. In the discrete implementation, Eq.~\eqref{eq:valley_chern} is evaluated as a restricted sum of FHS plaquette phases over the plaquettes contained in $P_K$. Throughout the optimization and main text, we use the non-negative magnitude $|C_v|\equiv |C_v^K|$. This convention removes the sign ambiguity between the two valleys while retaining sensitivity to Berry-curvature delocalization outside the chosen patch.

\subsection*{Optimization methodology and topological figure of merit}
\label{subsec:opt_and_fom}

Each unit cell is parameterized by
\[
\mathbf{x}=(l_1,l_2,\theta_1,\theta_2,N_1,N_2),
\]
where $l_1,l_2$ are normalized feature sizes, $\theta_1,\theta_2$ are in-plane rotations, and $N_1,N_2$ are polygon side counts for two symmetry-inequivalent inclusions. Continuous variables are bounded relative to the lattice constant, rotations are folded into the fundamental sector, and the side counts are restricted to the admissible set specified in the Supplementary Information.

We use a mixed-integer particle-swarm optimization (PSO). Continuous coordinates follow the standard velocity--position update with a decaying inertia weight, while the discrete variables are updated through copy/mutation steps and projected back to the admissible set. Each particle evaluation consists of unit-cell construction, PWE band calculation, FHS Berry-curvature evaluation, valley-Chern integration, and objective scoring.

The bulk topology--bandwidth objective is
\begin{equation}
\mathcal{T}(\mathbf{x})=
\left(\frac{\Delta f}{f_0}\right)^2 |C_v|,
\label{eq:fom}
\end{equation}
where $\Delta f$ is the minimum direct gap of the targeted valley-Hall band pair and $f_0$ is the mid-gap frequency. For numerical reporting and visualization, we use the scaled optimization score
$
\widetilde{\mathcal{T}}(\mathbf{x})
=
100\ \mathcal{T}(\mathbf{x})
$.

The factor of \(100\) is a constant numerical scaling introduced for visualization. It does not change the ordering of candidate structures, the location of the optimum, or the non-dominated Pareto set. Accordingly, the objective values shown in Fig.~\ref{fig:pareto_frontier_main} and Figs.~S8--S9 correspond to \(\widetilde{\mathcal{T}}\), whereas Eq.~\ref{eq:fom} defines the underlying unscaled figure of merit \(\mathcal{T}\).

The squared gap factor prevents the optimizer from favoring near-gap-closing geometries whose Berry curvature can be artificially enhanced by small interband denominators. Perturbatively,
\begin{equation}
\Omega_n(\mathbf{k})\sim -2\,\mathrm{Im}\sum_{m\ne n}
\frac{\langle u_n|\partial_{k_x}\hat{H}|u_m\rangle
\langle u_m|\partial_{k_y}\hat{H}|u_n\rangle}
{[\omega_n(\mathbf{k})-\omega_m(\mathbf{k})]^2},
\label{eq:berry_perturbative}
\end{equation}
so a Berry-curvature-only objective can spuriously favor small-gap designs. Equation~\eqref{eq:fom} is deliberately a bulk objective: it rewards spectral isolation and valley-localized Berry curvature, but it does not directly optimize edge-band coverage, group velocity, bend loss, or disorder-induced backscattering. These transport properties are evaluated after the bulk search.

\subsection*{Full-wave transport diagnostics}

To evaluate the selected bulk-optimized design at device scale, we construct domain-wall waveguides by interfacing the optimized lattice with its inverted partner using the defect-type interface shown in Fig.~\ref{fig:optimized}e. Two-dimensional Tidy3D FDTD simulations are used to compute straight and sharply bent waveguide transmission, field confinement, and time-averaged Poynting-vector flow. These simulations are interpreted as in-plane transport diagnostics within the same effective-index model used for optimization.

\subsection*{Disorder-averaged propagation-length analysis}

To obtain a quantitative robustness observable across multiple designs, we construct straight domain-wall waveguides of length $L=200a$ (lattice constant $a=423\ nm$) and apply independent Gaussian positional disorder of standard deviation $\sigma=0.01a$ to the photonic-crystal features. For each design, $N_{\mathrm r}=20$ disorder realizations are simulated at a design-specific carrier frequency selected from the high-transmission window of the FDTD spectrum of the corresponding waveguide. Following ensemble-decay methods used to quantify topological slow-light robustness \cite{Arregui2021}, we form cross-section-averaged normalized intensities from $|H_z|^2$ and, for consistency checks, from total electric-field intensity. The propagation length is extracted from
\begin{equation}
\left\langle \ln I_{\alpha,\mathrm{norm}}(x)\right\rangle
=A_\alpha-\frac{x}{\xi_\alpha},
\qquad \alpha\in\{H,E\},
\label{eq:xi_fit}
\end{equation}
after smoothing unit-cell-scale oscillations. Bootstrap resampling over disorder realizations gives the reported asymmetric confidence intervals. Because $\xi$ also depends on group velocity, modal confinement, source coupling, and interface termination, the resulting $|C_v|$--$\xi$ trend is interpreted as an empirical association under a common disorder protocol rather than as a universal causal law.

\section*{Results}

\subsection*{Topology--bandwidth landscape and Pareto tradeoff}

We first evaluate the mixed-integer PSO design population in terms of two bulk descriptors: the squared normalized gap $(\Delta f/f_0)^2$ and the absolute valley-Chern number $|C_v|$ of the lower band forming the target valley-Hall gap. These quantities enter the scalar objective in Eq.~\eqref{eq:fom}, but they are also plotted separately to expose the underlying multiobjective structure.

\begin{figure}[t]
 \centering

 \includegraphics[width=1\linewidth]{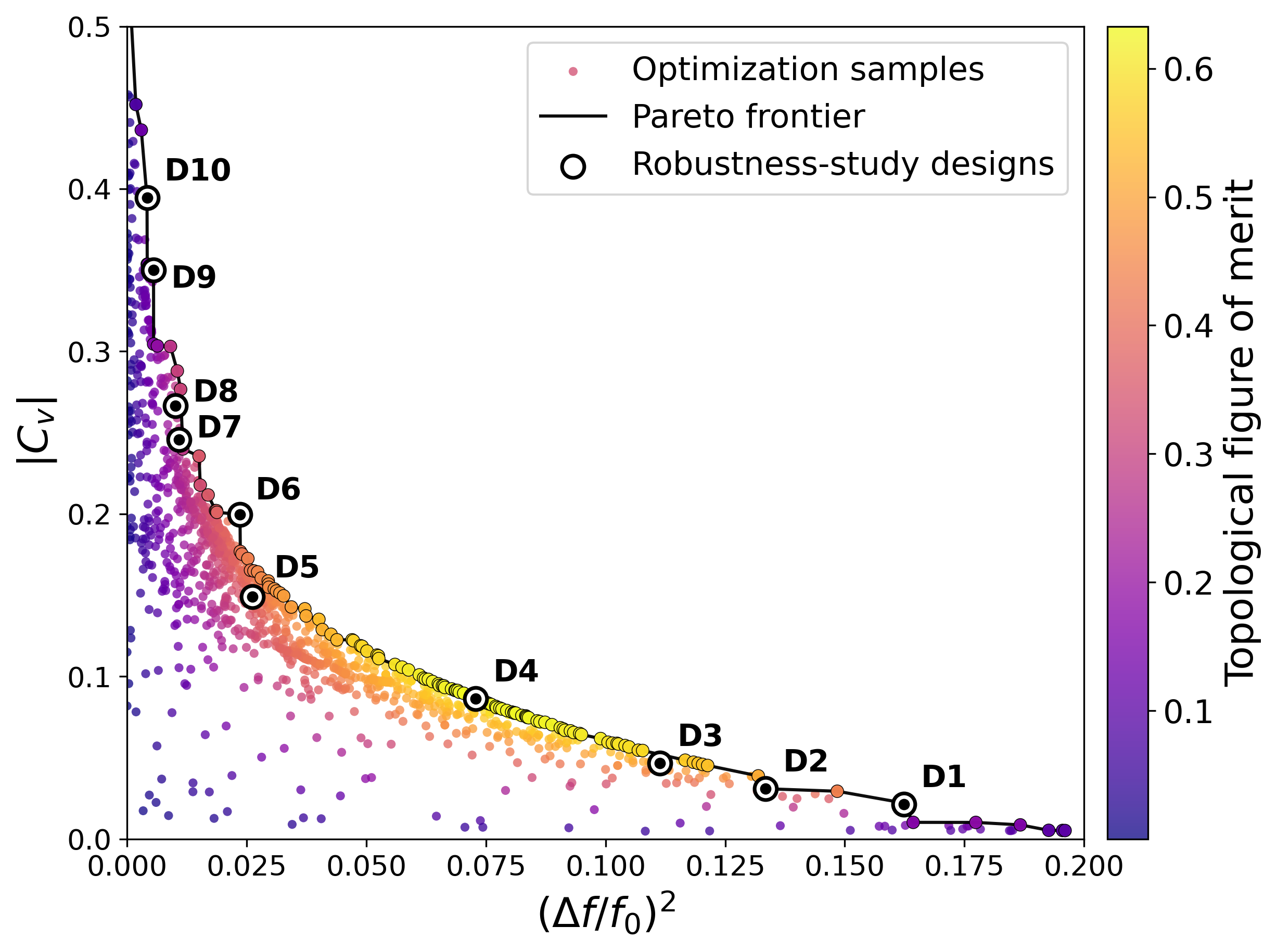}
 \caption{\textbf{Topology--bandwidth tradeoff and Pareto frontier obtained from mixed-integer PSO.}
 Each point corresponds to a candidate unit cell evaluated during optimization, plotted by the squared normalized bandgap $(\Delta f/f_0)^2$ and the absolute valley-Chern number $|C_v|$. The color encodes the scaled topological figure of merit $\widetilde{\mathcal{T}}= 100 \ (\Delta f/f_0)^2|C_v|$. The black curve marks the non-dominated Pareto frontier, showing that bandwidth and valley localization can be co-optimized but not simultaneously maximized within this design family. Open circles mark the ten designs D1--D10 selected for the disorder-averaged propagation-length analysis in Fig.~\ref{fig:cv_xi}. Design D4 is the optimized geometry used for the band, Berry-curvature, and straight/bent transport diagnostics in Figs.~\ref{fig:optimized} and \ref{fig:wg}.}
 \label{fig:pareto_frontier_main}
\end{figure}

Figure~\ref{fig:pareto_frontier_main} shows that the feasible set is not one-dimensional. Instead, the optimizer produces a cloud of solutions whose upper envelope forms a non-dominated Pareto frontier. Along this boundary, increasing the normalized gap generally reduces the patch-integrated valley-Chern magnitude because the Berry curvature spreads over a broader region of the Brillouin zone and becomes less confined to the valley patch. This directly visualizes the topology--bandwidth tradeoff targeted in this work.

\begin{figure*}[t]
 \centering
 \includegraphics[width=\textwidth]{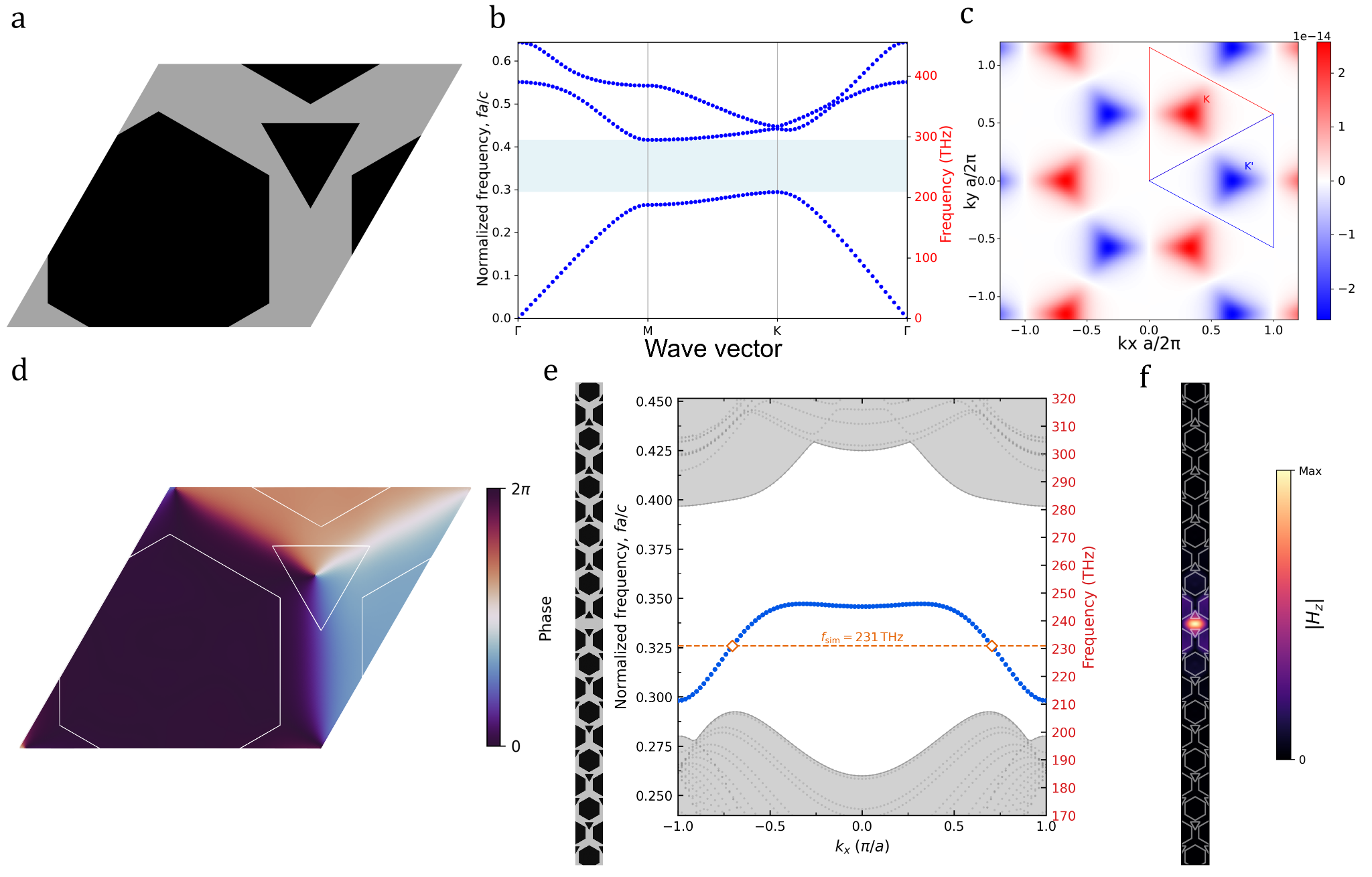}
 \caption{\textbf{Optimized unit cell and its properties.}
 \textbf{(a)} Unit-cell geometry obtained by mixed-integer PSO; inversion symmetry is broken by two symmetry-inequivalent inclusions.
 \textbf{(b)} Photonic band structure along $\Gamma\! -\! M\! -\! K\! -\! \Gamma$ with the optimized bulk gap highlighted.
 \textbf{(c)} Berry curvature $\Omega(\mathbf{k})$ over the Brillouin zone. Red and blue outlines show the fixed, non-overlapping $K$- and $K'$-valley integration patches used to compute the signed valley-Chern numbers. The optimization uses the magnitude $|C_v|=|C_v^K|$.
 \textbf{(d)} Representative real-space phase profile of a valley Bloch eigenmode for the optimized unit cell.
 \textbf{(e)} Projected domain-wall band structure calculated using COMSOL Multiphysics, with the corresponding supercell geometry shown alongside. The gray regions and faint gray points denote the projected bulk continua and constituent bulk eigenfrequencies, while the blue points identify the tracked in-gap edge-state branch localized at the intended domain wall. The horizontal orange dashed line marks the carrier frequency used in the finite-device FDTD simulations, \(f_{\mathrm{sim}}\equiv f_{\mathrm c}=231\,\mathrm{THz}\), and the open orange markers indicate its two reciprocal intersections with the edge-state dispersion.
 \textbf{(f)} Representative $|H_z|$ field profile of the domain-wall mode indicated in panel (e), showing confinement to the interface.}
 \label{fig:optimized}
\end{figure*}

\subsection*{Optimized unit cell, valley gap, and topological characteristics}

The selected optimized design is shown in Fig.~\ref{fig:optimized}a. The two symmetry-inequivalent inclusions break inversion symmetry and open a valley-Hall gap between the target bands. The primitive-cell band structure in Fig.~\ref{fig:optimized}b defines the bulk gap $\Delta f$ and mid-gap frequency $f_0$ used in Eq.~\eqref{eq:fom}. The Berry-curvature map in Fig.~\ref{fig:optimized}c shows opposite-sign localization near $K$ and $K'$ and explicitly marks the integration patches used to compute the signed valley-Chern numbers. This graphical patch definition is important because the non-quantized value of $|C_v|$ depends on how much curvature remains inside the valley sector.

The projected domain-wall spectrum in Fig.~\ref{fig:optimized}e is a distinct eigenvalue problem from the primitive-cell band structure in Fig.~\ref{fig:optimized}b. The former uses a supercell containing two inverted domains and a specific defect-type interface, so its localized branch dispersion depends on the domain-wall termination. Consequently, the projected interface-mode window is not expected to coincide with the entire primitive-cell direct gap. The usable waveguide window is instead the frequency range where a localized interface branch with suitable group velocity exists and can be excited in the finite FDTD structure. The field profiles in Fig.~\ref{fig:optimized}f confirm confinement of the representative domain-wall modes.

This distinction is central to the interpretation of the figure of merit. Equation~\eqref{eq:fom} selects a bulk topology--bandwidth compromise; it does not guarantee that an edge branch spans every frequency in the primitive-cell gap or that the branch has uniformly large group velocity. These interface properties are therefore evaluated separately through the transport diagnostics below.

\begin{figure*}[t]
 \centering
 \includegraphics[width=\textwidth]{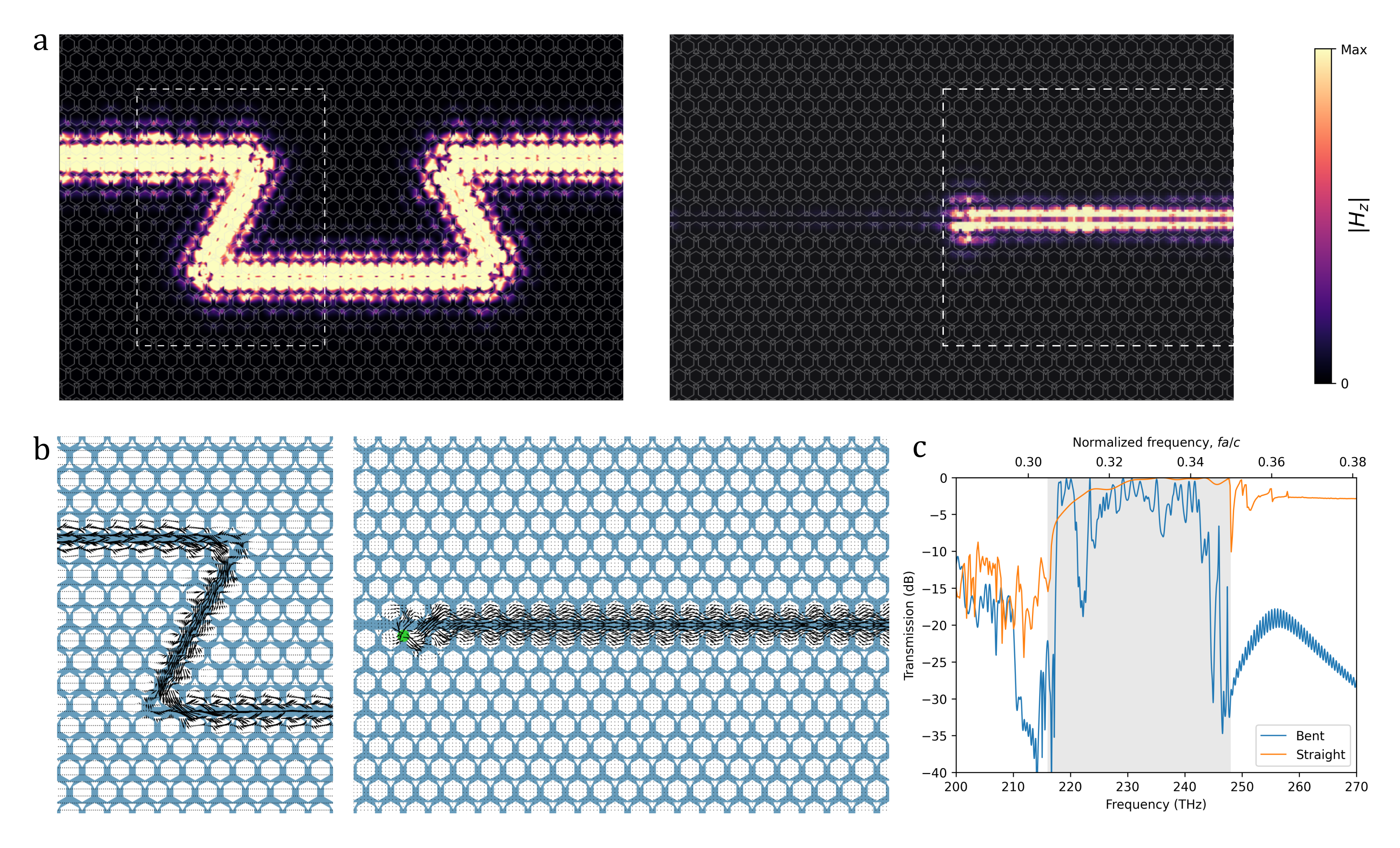}
 \caption{\textbf{Full-wave transport diagnostics of the selected optimized VPC design.}
 \textbf{(a)} Steady-state field intensity $|H_z|$ for a compact sharp-bend domain-wall route and a straight domain-wall route built from the optimized unit cell and its inverted partner.
 \textbf{(b)} Poynting-vector maps showing energy flow along the interface for the bent and straight geometries in the boxed regions of panel (a).
 \textbf{(c)} Transmission spectra for the bent and straight waveguides. The gray shaded region denotes the usable interface-mode window within the two-dimensional effective-index model.}
 \label{fig:wg}
\end{figure*}

\subsection*{Device-scale waveguiding through straight and bent interfaces}

We next evaluate the selected geometry in finite-domain FDTD simulations. Using the same defect-type interface as in Fig.~\ref{fig:optimized}e, we construct a straight domain-wall waveguide and a compact sharp-bend route containing multiple corners. Figure~\ref{fig:wg}a shows that the field remains confined near the interface in both geometries, while Fig.~\ref{fig:wg}b shows that the time-averaged energy flow follows the domain-wall path.

The spectra in Fig.~\ref{fig:wg}c are interpreted as transport diagnostics rather than as proof of topological protection. Within the usable interface-mode window, the bent and straight structures both transmit through the finite two-dimensional domain, but the bent spectrum is reduced relative to the straight case. This reduction indicates residual bend mismatch and scattering, and confirms that the bulk objective does not by itself eliminate all interface loss. Near the band edges, degradation is attributed within the present model to coupling to bulk photonic-crystal states, mode mismatch, and finite-structure scattering. Vertical leakage and substrate radiation are outside the two-dimensional effective-index approximation.

\begin{figure*}[t]
    \centering
    \includegraphics[width=\textwidth]{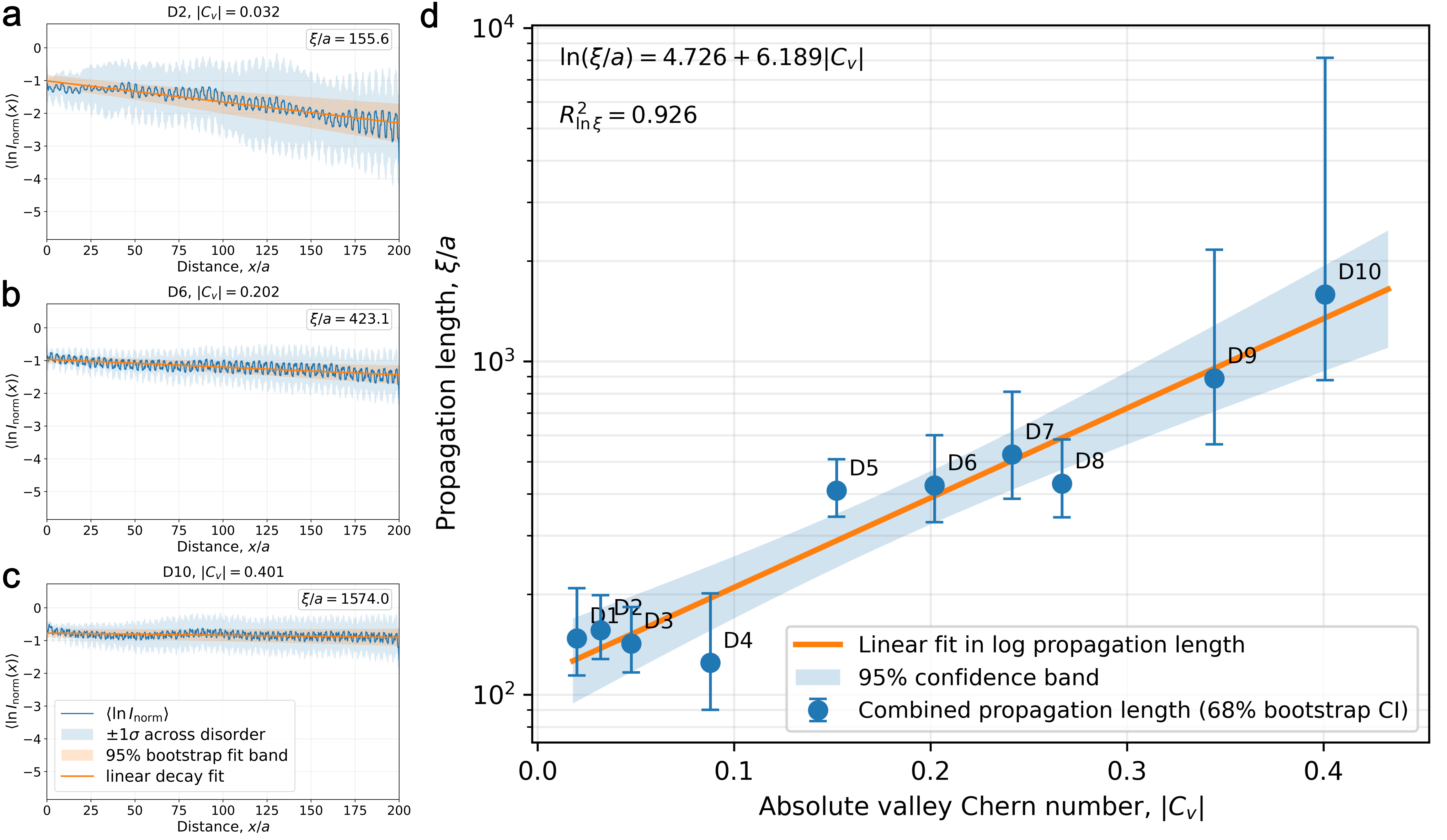}
    \caption{\textbf{Empirical relation between the valley-Chern number and disorder-limited propagation length.}
    \textbf{(a--c)} Representative propagation-length extraction for designs D2, D6, and D10, spanning low, intermediate, and high values of $|C_v|$. The blue curves show the ensemble-averaged normalized logarithmic intensity, $\langle \ln I_{\mathrm{norm}}(x)\rangle$, along a $200a$ domain-wall waveguide with positional disorder $\sigma=0.01a$. The light-blue regions denote the $\pm 1\sigma$ variation across disorder realizations, the orange lines are fits to $\langle \ln I_{\mathrm{norm}}(x)\rangle=A-x/\xi$, and the orange shaded regions denote 95\% bootstrap confidence intervals. The extracted propagation lengths are $\xi/a=155.6$, $423.1$, and $1574.0$ for D2, D6, and D10, respectively.
    \textbf{(d)} Combined propagation length $\xi/a$ versus $|C_v|$ for the design ensemble. Vertical error bars indicate 68\% bootstrap confidence intervals. The orange line is a log-linear regression, $\ln(\xi/a)=4.726+6.189|C_v|$, with $R_{\ln\xi}^2=0.926$; the blue shaded region denotes the 95\% confidence interval of the fitted mean relation. The trend is interpreted as a positive empirical association within the investigated VPC design family, not as a universal one-parameter scaling law.}
    \label{fig:cv_xi}
\end{figure*}

\subsection*{Disorder-limited propagation length across the design ensemble}

Figure~\ref{fig:cv_xi} provides the quantitative transport metric plotted against $|C_v|$. Low-$|C_v|$ designs in the present ensemble have propagation lengths of order $10^2a$, whereas designs with larger $|C_v|$ generally sustain propagation over several hundred lattice constants; the largest-$|C_v|$ design reaches the $10^3a$ scale, although with a broad confidence interval because the fitted decay becomes weak over a finite $200a$ simulation length.

The normalized-gap factor in Eq.~\eqref{eq:fom} is intended to reward bandwidth rather than directly suppress scattering. Figure~\ref{fig:cv_xi} therefore specifically tests the physical relevance of the $|C_v|$ factor in the bulk objective. The result should not be interpreted as evidence that $\mathcal{T}$ is a universal predictor of edge transport. Rather, under a common disorder protocol, stronger valley-localized Berry flux is associated with longer ensemble-averaged propagation. Interface dispersion, group velocity, source coupling, local mode matching, and termination remain additional variables, and direct backscattering or propagation length can be incorporated as a third objective in future device-level optimization.

\section*{Discussion}

This work frames inverse design in valley photonic crystals as an explicit topology--bandwidth optimization problem. The central tradeoff is between a large normalized valley gap and the degree to which the Berry curvature remains localized near a valley so that $|C_v|$ remains close to its small-gap value. In broad-gap regimes, the patch-integrated valley-Chern number is approximate because Berry curvature can spread across the Brillouin zone. The Pareto frontier in Fig.~\ref{fig:pareto_frontier_main} makes this compromise explicit.

The disorder-ensemble result in Fig.~\ref{fig:cv_xi} adds a device-level interpretation to this bulk tradeoff. Within the tested design family and under the same $\sigma=0.01a$ disorder protocol, larger $|C_v|$ is associated with longer propagation length. This supports the topology-aware character of the bulk objective, but it does not imply that $C_v$ alone controls scattering. In realistic photonic-crystal interfaces, residual loss depends on intervalley Fourier components of the disorder, group velocity and density of states, modal confinement, source coupling, and local interface termination.

An important limitation is that bend transmission is not a unique diagnostic of topology. Recent work has shown that high transmission through sharp photonic-crystal bends can also arise from interface-specific mode matching, local field chirality, and polarization-singularity structure \cite{Dai2024,Dai2025}. We therefore distinguish between bulk topological descriptors, such as Berry-curvature localization and $C_v$, and device-level observables, such as straight-versus-bent transmission and disorder-limited propagation length. The present scalar FOM is intentionally a bulk topology--bandwidth objective, not a direct minimization of edge-state loss. This distinguishes the present question from inverse-design strategies that explicitly optimize disorder-induced backscattering at fixed group index \cite{Thompson2026Backscattering}.

The distribution of optimized geometries also provides physical insight. High-FOM solutions preferentially occur for polygonal features compatible with the threefold symmetry structure of the hexagonal valley lattice. This is natural because valley-Hall physics is controlled by the $K/K'$ valleys, whose mode structure is constrained by local rotational symmetry. Geometries that open an inversion-breaking mass while preserving valley separation tend to retain stronger valley-contrasting Berry curvature, whereas less symmetry-compatible perturbations can delocalize Berry curvature or increase intervalley mixing.

A natural extension is to add $\xi$, bend excess loss, edge-band coverage, minimum group velocity, or a large-momentum intervalley-overlap functional as an additional objective or constraint. Such a formulation would retain the physically interpretable Pareto map developed here while directly optimizing a selected device-level robustness metric.

Table~\ref{tab:wg_comparison} gives a representative operating-band comparison against selected topological photonic-crystal waveguides from the literature. The relative bandwidth is defined as
\begin{equation}
\mathrm{RBW}=\frac{f_+-f_-}{f_0}\times 100\%,\qquad f_0=\frac{f_++f_-}{2},
\label{eq:rbw}
\end{equation}
where $f_-$ and $f_+$ denote the lower and upper edges of the reported operating window. For wavelength-domain entries, we compute the equivalent frequency-domain fractional bandwidth using $\mathrm{RBW}=2(\lambda_+-\lambda_-)/(\lambda_++\lambda_-)\times100\%$. The table is representative rather than exhaustive, and direct comparison of robustness requires caution because different works use different platforms, dimensionalities, interfaces, and transport metrics.

\begin{table*}[t]
\centering
\caption{\textbf{Representative operating-band comparison for selected topological photonic-crystal waveguides.}
RBW denotes the relative bandwidth defined in Eq.~\eqref{eq:rbw}. The comparison is representative rather than exhaustive, and robustness metrics are not directly equivalent across works because the platforms, dimensionalities, and interface geometries differ.}
\label{tab:wg_comparison}
\small
\setlength{\tabcolsep}{4pt}
\renewcommand{\arraystretch}{1.15}
\begin{tabular}{@{}l l l c c l@{}}
\toprule
Work & Platform & Topological class & Edge-mode band & RBW (\%) & Notes \\
\midrule
\tblnote[2.6cm]{\textbf{This work}} &
\tblnote[2.1cm]{2D effective-index Si VPC model} &
\tblnote[2.2cm]{Valley Hall} &
$213$--$249~\mathrm{THz}$ &
15.6 &
\tblnote[5.6cm]{Straight and bent domain-wall transmission and disorder-limited propagation length evaluated in a 2D effective-index model.} \\
\tblnote[2.6cm]{He et al. \cite{He2019}} &
\tblnote[2.1cm]{SOI VPC slab} &
\tblnote[2.2cm]{Valley Hall} &
$1360$--$1492~\mathrm{nm}$ &
$\sim 9.3$ &
\tblnote[5.6cm]{TE-like VPC bandgap; RBW computed on the frequency scale.} \\
\tblnote[2.6cm]{Shalaev et al. \cite{Shalaev2019}} &
\tblnote[2.1cm]{Silicon VPC} &
\tblnote[2.2cm]{Valley Hall} &
183--193~THz &
5.3 &
\tblnote[5.6cm]{Representative telecom-scale valley-Hall transport.} \\
\tblnote[2.6cm]{Yang et al. \cite{Yang2020}} &
\tblnote[2.1cm]{Silicon VPC} &
\tblnote[2.2cm]{Valley Hall} &
320--350~GHz &
9.0 &
\tblnote[5.6cm]{THz valley-Hall waveguide demonstration.} \\
\tblnote[2.6cm]{Webber et al. \cite{Webber2021}} &
\tblnote[2.1cm]{THz VPC waveguide} &
\tblnote[2.2cm]{Valley Hall} &
320--350~GHz &
9.0 &
\tblnote[5.6cm]{} \\
\tblnote[2.6cm]{Liu et al. \cite{Liu2022}} &
\tblnote[2.1cm]{THz Si VPC switch} &
\tblnote[2.2cm]{Valley Hall} &
266--290~GHz &
8.6 &
\tblnote[5.6cm]{Experimentally demonstrated THz topological waveguide switch.} \\
\tblnote[2.6cm]{Tan et al. \cite{Tan:22}} &
\tblnote[2.1cm]{THz Si VPC} &
\tblnote[2.2cm]{Valley Hall} &
405--488~GHz &
18.6 &
\tblnote[5.6cm]{Broadband operation enabled by composite interface engineering.} \\
\tblnote[2.6cm]{Barik et al. \cite{doi:10.1126/science.aaq0327}} &
\tblnote[2.1cm]{Wu--Hu photonic crystal} &
\tblnote[2.2cm]{Spin Hall} &
313--333~THz &
6.2 &
\tblnote[5.6cm]{} \\
\bottomrule
\end{tabular}
\end{table*}

\section*{Conclusion}

We introduced a mixed-integer inverse-design framework for valley photonic crystals that makes the topology--bandwidth tension explicit and quantitatively navigable. By jointly evaluating the normalized valley bandgap fraction $\Delta f/f_0$ and the patch-integrated valley-Chern number $|C_v|$, and by organizing the resulting design population through non-dominated sorting, we obtain a Pareto frontier showing which combinations of bandwidth and valley localization are attainable within the chosen geometry class.

The disorder-ensemble calculations provide an independent transport-level test of the valley-Chern factor. Under a common positional-disorder protocol with $\sigma=0.01a$, the extracted propagation length $\xi$ exhibits a positive empirical association with $|C_v|$ across the selected design ensemble. This supports the physical relevance of the valley-Chern component while preserving the more limited and defensible claim that the FOM selects a bulk topology--bandwidth compromise rather than guaranteeing optimal edge transport.

Overall, the proposed Pareto-driven methodology provides a computationally efficient and physically interpretable route for topology-aware photonic-crystal design within a two-dimensional effective-index framework. The same framework can be extended to include additional objectives or constraints, including disorder sensitivity, interface-mode matching, fabrication tolerances, and ultimately full three-dimensional slab radiation losses, so that performance metrics, topological descriptors, and device limitations can be optimized on equal footing.


\section*{Funding Sources}
Anshuman Kumar acknowledges funding support from the National Quantum Mission, an initiative of the Department of Science and Technology, Government of India. Anshuman Kumar also acknowledges funding from ANRF via grant number SPR/2023/000175. Abhishek Kumar acknowledges support from ANRF through the Prime Minister Early Career Research Grant under grant number ANRF/ECRG/2024/001205/PMS.

\bibliographystyle{achemso}
\bibliography{references.bib}









\end{document}